\documentclass[OptSoft]{jtcam_preprint}

\usepackage{hyperref}
\usepackage{subcaption}
\usepackage{upgreek}
\usepackage{tikz}
\usepackage{pgfplots}
\usepackage{bm}

\title{How to introduce an initial crack in phase field simulations to accurately predict the  linear elastic fracture propagation threshold?}
\runningtitle{Introducing initial cracks in phase field fracture simulations}

\author[1]{\orcid{0000-0002-6305-7309} Flavien Loiseau}
\author[1]{\orcid{0000-0003-0081-2650} Véronique Lazarus}
\runningauthor{F. Loiseau et al.}
\affil[1]{IMSIA, ENSTA, EDF, CNRS, Palaiseau, 91120, France}


\newcommand{\xx}{\bm{x}}
\newcommand{\dx}{\,\mathrm{d}\xx}
\newcommand{\uu}{\bm{u}}
\newcommand{\uimp}{\uu_{\mathrm{imp}}}
\newcommand{\uimpbar}{\widebar{\uu}_{\mathrm{imp}}}
\newcommand{\fv}{\bm{f}}
\newcommand{\ts}{\bm{t}}
\newcommand{\nn}{\bm{n}}

\newcommand{\ela}{\mathbf{E}}

\newcommand{\eps}{\bm{\upvarepsilon}}
\newcommand{\sig}{\bm{\upsigma}}

\definecolor{tabblue}{HTML}{1f77b4}
\definecolor{taborange}{HTML}{ff7f0e}
\definecolor{tabgreen}{HTML}{2ca02c}
\definecolor{tabred}{HTML}{d62728}
\definecolor{tabpurple}{HTML}{9467bd}
\definecolor{tabbrown}{HTML}{8c564b}
\definecolor{tabpink}{HTML}{e377c2}
\definecolor{tabgray}{HTML}{7f7f7f}
\definecolor{tabolive}{HTML}{bcbd22}
\definecolor{tabcyan}{HTML}{17becf}
\definecolor{semiana}{HTML}{d62728}       
\definecolor{uimp1}{HTML}{2ca02c}         
\definecolor{uimp2}{HTML}{277625}
\definecolor{uimp3}{HTML}{204e1c}
\definecolor{uimp4}{HTML}{162913}
\definecolor{cmsi1}{HTML}{1f77b4}       
\definecolor{cmsi2}{HTML}{205783}
\definecolor{cmsi3}{HTML}{1c3a56}
\definecolor{cmsi4}{HTML}{14202d}
\definecolor{cndi1}{HTML}{ff7f0e}       
\definecolor{cndi2}{HTML}{ba5e15}
\definecolor{cndi3}{HTML}{794014}
\definecolor{cndi4}{HTML}{3e230f}
\definecolor{geometric}{HTML}{bcbd22}         
\definecolor{embedded}{HTML}{17becf}          

\begin{document}

\def\version{Preprint, version of \today.}

\begin{abstract}
Variational phase field fracture models are now widely used to simulate crack propagation in structures.
A critical aspect of these simulations is the correct determination of the propagation threshold of pre-existing cracks, as it highly relies on how the initial cracks are implemented.
While prior studies briefly discuss initial crack implementation techniques, we present here a systematic investigation.
Various techniques to introduce initial cracks in phase field fracture simulations are tested, from the crack explicit meshing to the replacement by a fully damaged phase field, including different variants for the boundary conditions.
Our focus here is on phase field models aiming to approximate, in the $\Gamma$-convergence limit, Griffith quasi-static propagation in the framework of Linear Elastic Fracture Mechanics. 
Therefore, a sharp crack model from classic linear elastic fracture mechanics based on Griffith criterion is the reference in this work.
To assess the different techniques to introduce initial cracks, we rely on path-following methods to compute the sharp crack and the phase field smeared crack solutions.
The underlying idea is that path-following ensures staying at equilibrium at each instant so that any difference between phase field and sharp crack models can be attributed to numerical artifacts.  
Thus, by comparing the results from both models, we can provide practical recommendations for reliably incorporating initial cracks in phase field fracture simulations.
The comparison shows that an improper initial crack implementation often requires the smeared crack to transition to a one-element-wide phase band to adequately represent a displacement jump along a crack.
This transition increases the energy required to propagate the crack, leading to a significant overshoot in the force-displacement response.
The take-home message is that to predict the propagation threshold accurately and avoid artificial toughening; the crack must be initialized either setting the phase field to its damage state over a one-element-wide band or meshing the crack explicitly as a one-element-wide slit and imposing the fully cracked state on the crack surface.
\end{abstract}


\section{Introduction}

Phase field fracture models based on the variational approach to fracture, proposed by \textcite{francfort_revisiting_1998} and implemented by \textcite{bourdin_numerical_2000}, have become increasingly popular in the fracture mechanics community.
These models are based on \textcite{griffith_phenomena_1920} idea that the mechanical state of a structure is governed by the balance between the elastic energy and the fracture surface energy.
This idea can also be formalized as the minimization of the potential energy among any possible crack extension, that is any sharp discontinuity of the displacement field \parencite{francfort_revisiting_1998}.
As the crack surface is not known \emph{a priori}, performing minimization on this set of sharp cracks is generally not possible in practice. 
The phase field fracture models replace the sharp crack with a smeared crack represented by a phase field and use a regularization of the elastic and fracture energies based on the work of \textcite{ambrosio_approximation_1990, ambrosio_approximation_1992}.
For variational phase field models, the regularized functional has been shown to $\Gamma$-converge towards its sharp counterpart \parencite{bellettini_discrete_1994, bourdin_image_1999, giacomini_discontinuous_2003, chambolle_approximation_2004}.
This property means that the results obtained using phase field fracture models, with a small-enough regularization length and sufficient refinement, converges towards the sharp crack model using Griffith theory.


Recent studies have highlighted the limitations of classical variational phase field models in describing crack nucleation from \emph{ab-nihilo} \parencite{kumar_revisiting_2020, bazant_critical_2022, lopez-pamies_classical_2024}.
However, these models effectively handle crack propagation problems.
To deal with the failure of a structure using a variational phase field model, it is therefore necessary to introduce an initial crack, as in standard Linear Elastic Fracture Mechanics (LEFM).
But, how initial cracks are introduced in phase field simulations can significantly bias the observed mechanical response, in particular the propagation threshold.
For instance, the work of \textcite{singh_fracture-controlled_2016} shows that an overshoot of the force peak occurs in the force-displacement response of the structure without a specific treatment applied to the crack tip.
Similarly, \textcite{kristensen_assessment_2021} also observed an artificial increase of the critical energy release rate when introducing crack geometrically in the simulation domain.
It underscores the importance of having a clear and reliable method to introduce initial cracks as it affects when and how initial cracks propagates.

The problem of implementing initial cracks in phase field fracture simulation is discussed in the literature, but in a scattered manner and often as a side study.
A first technique, used by \textcite{klinsmann_assessment_2015,singh_fracture-controlled_2016,kristensen_assessment_2021}, among others, consists in explicitly embedding the initial cracks into the mesh by duplicating nodes, resulting in an infinitely sharp crack.
This technique is largely employed in Finite Element fracture simulations \parencite[Section 12.4.]{anderson_fracture_2017}.
However, its application to phase field fracture simulations raises several questions.
Indeed, one challenge is determining the appropriate boundary condition to impose on the initial crack.
For the displacement field, stress-free boundary conditions has to be used for unloaded crack.
However, different boundary conditions in crack phase coexist in the literature.
The phase field variable can be: (a) not set explicitly (zero Neumann boundary condition), (b) set to a fully cracked state at the crack tip, or (c) set to a fully cracked state on the whole crack.
Similarly, \textcite{tanne_crack_2018} proposed to replace the initial crack with a slightly open sharp V-notch.
A second technique consists in including the initial crack in the initial phase field.
With this implementation technique, the initial crack is not meshed explicitly, but the crack is introduced instead by setting the initial phase field.
More specifically, the crack phase is set to a fully cracked state along the initial crack \parencite{nguyen_phase_2015,makvandi_phase-field_2019, lo_phase-field_2019, yoshioka_crack_2020, kristensen_assessment_2021}.
A third technique can be applied in phase field models where a history variable governs the crack phase evolution \parencite{miehe_phase_2010}.
The initial crack phase field is imposed indirectly through the history variable.
This method has been proposed by \textcite{borden_phase-field_2012} and employed by \textcite{klinsmann_assessment_2015, pham_experimental_2017, liu_investigation_2020}, among others.
As this method only applies to the phase field models using a history variable, this approach will only be briefly discussed  in this work.
For all those methods, a crucial factor is the initial crack thickness.
This aspect is often unspecified in the literature (or even overlooked in some cases).
However, it can significantly impact the numerical predictions of phase field fracture simulations as it will be shown here.

In this article, we explore systematically the impact of various initial crack implementation techniques on the mechanical response of structures.
The sharp crack model will act as our reference as the phase field smeared crack model $\Gamma$-converges toward it.
Our aim is to determine whether certain crack implementation technique introduce a bias into the mechanical results, in particular the critical load, and why.
The comparison of the phase field fracture with the reference sharp crack model enables us to identify the most reliable technique to introduce initial cracks in phase field simulations.
The findings will help users of the phase field method to choose the most accurate approach.
Focus will be placed on the quasi-static propagation that satisfies Griffith criterion at each moment using a path-following method.
For each crack growth increment, the corresponding critical load will be determined explicitly in sharp crack models or through a path-following constraint in phase field simulations.
Following the quasi-static equilibrium path during crack propagation enables us to highlight the errors induced by an improper crack initialization irrespective of any dynamical effect.

This paper is organized as follows.
\Cref{sec:variational_approach_to_fracture} introduces the sharp crack model through the Griffith theory, and its formalization in the variational approach to fracture.
The smeared crack model (\emph{i.e.,} the phase field fracture model) is then presented as a regularization of the variational approach to fracture.
\Cref{sec:path-following} defines the quasi-static equilibrium path and explains how it is obtained, analytically and numerically, for the sharp and smeared crack.
\Cref{sec:crack_initialization_techniques} presents various techniques to introduce initial cracks in phase field fracture simulation.
\Cref{sec:comparison_crack_initialization_techniques} compares eight different techniques to implement initial cracks in phase field simulations on a simple problem (Single Edge Notched Tension test).
The results obtained in this section enable us to discuss the different techniques and determinate which ones are unbiased.
The whole study has been done for straight cracks and phase field models without history variables.
\Cref{sec:discussion} briefly discusses  how the results can be extended beyond.
Finally, \cref{sec:conclusion} concludes the study and provides practical recommendations to introduce initial cracks in phase field fracture simulations.

\section{From Griffith theory to variational phase field models}
\label{sec:variational_approach_to_fracture}

Starting from the original idea of \textcite{griffith_phenomena_1920}, we recall the variational approach to fracture proposed by \textcite{francfort_revisiting_1998}.
This section covers the theoretical grounds of this work with an emphasis on the link between Griffith theory and variational phase field models for fracture.
\textcite{bourdin_variational_2008} provides a more detailed discussion on this subject.

\subsection{Griffith theory}

For a static elastic problem, the displacement field of an elastic body submitted to external loads minimizes its potential energy $\mathcal{P} = \mathcal{E} - \mathcal{W}_{\mathrm{ext}}$, where $\mathcal{E}$ is the body elastic energy and $\mathcal{W}_{\mathrm{ext}}$ is the external work. In a 2D setting, these quantities are defined by unit thickness.
\textcite{griffith_phenomena_1920} proposed to extend this minimization principle to study the rupture of a solid with a crack $\Gamma$ of length $a$.
To this aim, \textcite{griffith_phenomena_1920} introduced the surface energy $G_c$, also called the fracture energy, defined as the energy required to generate new free surfaces\footnote{While considered as a reversible energy in \textcite{griffith_phenomena_1920}, we consider here that $G_c$ includes all the energy dissipated during propagation.}.
From this statement, the state of the solid at equilibrium must minimize the energy $\mathcal{P} + G_c a$.
Deriving this energy with respect to the crack length $a$ leads to the definition of the energy release rate
\begin{equation}
G = - \frac{\partial \mathcal{P}}{\partial a} \text{ at constant load.}
\label{eq:defG}
\end{equation}
To ensure the stability of a given crack equilibrium state, $G \leq G_c$ must be imposed, as $G > G_c$ implies the existence of an excess of energy, hence dynamical propagation.
Moreover, introducing an irreversibility constraint and deriving this energy with respect to the crack surface give the following quasi-static crack  propagation conditions:
\begin{equation}
  (G - G_c) \dot{a} = 0, \quad G \leq G_c, \quad
  \dot{a} \geq 0.
  \label{eq:griffith}
\end{equation}
It means that two distinct cases can occur.
If $G < G_c$, then $\dot{a} = 0$, which means that the crack does not propagate.
  Conversely, if $G = G_c$, then $\dot{a} \geq 0$, indicating that the crack propagation becomes possible.

\subsection{Variational approach to fracture}

\textcite{francfort_revisiting_1998} extended this work by formalizing this idea and writing Griffith theory in a variational framework as follows.

Let us consider an elastic body $\Omega$ with an initial crack $\Gamma_0$.
Body forces $\fv$ are prescribed in the bulk. 
The body boundary, denoted $\partial \Omega$, includes two distinct subsets: $\partial \Omega_{\uu}$ and $\partial \Omega_{\ts}$, where displacement $\uu$, respectively, surface forces are prescribed
\begin{equation}
    \uu=\uimp \text{ on }\partial \Omega_{\uu}
    \text{ and }
    \sig \cdot \nn = \ts \text{ on } \partial \Omega_{\ts}.
\end{equation}
Crack surfaces are assumed to be free of any external load.
Denote $\mathcal{U}_{\Gamma}$ the space of admissible displacements defined as
\begin{equation}
  \mathcal{U}_{\Gamma} = \left\{ \uu \in \mathrm{H}^1(\Omega \setminus \Gamma) \,\vert\, \forall \xx \in \partial \Omega_{\uu}, \uu(\xx) = \uimp(\xx) \right\},
\end{equation}
where  $\mathrm{H}^1(\Omega \setminus \Gamma)$ is a classic Sobolev space.
In this case, the elastic energy and the external work are
\begin{equation}
  \begin{split}
    &\mathcal{E}(\uu, \Gamma) = \int_{\Omega \setminus \Gamma}\frac{1}{2} \eps(\uu) : \ela : \eps(\uu) \dx
    , \\
    &\mathcal{W}_{\mathrm{ext}} (\uu) = \int_{\Omega} \fv \cdot \uu \dx + \int_{\partial \Omega_{\ts}} \ts \cdot \uu \dx
    , \\
  \end{split}
\end{equation}
where
  $\eps(\uu) = \frac{1}{2} (\nabla \uu + \nabla^{\mathrm{T}} \uu)$ is the strain,
  $\ela$ is the elasticity tensor.
Griffith theory, in \cref{eq:griffith}, can be rewritten as the minimization problem
\begin{equation}
  \label{eq:griffith_principle}
  (\uu, \Gamma) = \arg\min_{\substack{\uu' \in \mathcal{U}_{\Gamma} \\ \Gamma' \supseteq \Gamma_0}}
  \mathcal{E}(\uu', \Gamma') - \mathcal{W}_{\mathrm{ext}}(\uu') + \int_{\Gamma'} G_c \dx.
\end{equation}
The constraint $\Gamma' \supseteq \Gamma_0$ enforces the crack growth (\emph{i.e.}, the cracking irreversibility preventing cracks from healing).
Note that in this framework, a crack is represented by a stress-free surface that can propagate in the domain $\Omega$.
It induces a sharp discontinuity in the displacement field over which a displacement jump can occur.

\subsection{Regularization as variational phase field model}
As the crack corresponds to a discontinuity in the elastic body, the minimization problem in \cref{eq:griffith_principle} is a free-discontinuity problem, which is a well-studied problem \parencite{ambrosio_functions_2000}.
In the context of image segmentation, free-discontinuity problems have been regularized so that sharp discontinuities are replaced by smeared band of non-vanishing width \parencite{mumford_optimal_1989, ambrosio_approximation_1992}.
\textcite{francfort_revisiting_1998} proposed to apply the same regularization to the minimization problem in \cref{eq:griffith_principle}.
Applying the regularization replace the explicit description of the discontinuity surface $\Gamma$ with a new field: the (crack) phase field, denoted $\alpha(\xx)$.
By convention, we consider that the material at coordinate $\xx$ is uncracked if $\alpha(\xx)=0$ and fully crack if $\alpha(\xx) = 1$.
The minimization problem in \cref{eq:griffith_principle} governing the state of the domain becomes
\begin{equation}
  \label{eq:phase field_minimization}
  (\uu, \alpha) = \arg\min_{\substack{\uu \in \mathcal{U} \\ \alpha \in \mathcal{A}}}
  \mathcal{E}(\uu, \alpha) + \mathcal{D}(\alpha) - \mathcal{W}_{\mathrm{ext}} (\uu)
\end{equation}
where $\mathcal{U}$ and $\mathcal{A}$ are respectively the admissibility space of the displacement field $\uu$ and of the crack phase field $\alpha$.
The irreversibility of cracking is imposed through an irreversibility constraint on the phase field.
Given an initial phase field $\alpha_0(\xx)$, the phase field can stay constant or only grow, leading to the admissibility space
\begin{equation}
  \mathcal{A} = \left\{ \alpha \in \mathrm{H}^1(\Omega) \,\vert\, \forall \xx \in \Omega, \alpha(\xx) \geq \alpha_0(\xx) \right\}.
\end{equation}
The term $\mathcal{D}(\alpha)$ is the regularization of the dissipation integral over the (unknown) crack surface into a domain integral.
The different energies are defined as
\begin{equation}
  \begin{split}
    &\mathcal{E}(\uu, \alpha) = \int_\Omega \frac{1}{2} a(\alpha) \eps(\uu) : \ela : \eps(\uu) \dx
    , \\
    &\mathcal{D}(\alpha) = \frac{G_c}{c_w} \int_\Omega \frac{w(\alpha)}{\ell} + \ell \nabla \alpha \cdot \nabla \alpha \dx
    . \\
  \end{split}
\end{equation}
In addition to the classic elastic parameters and the critical energy release rate $G_c$, the regularization introduces a new parameter: the regularization length $\ell$, to which the thickness of the crack phase profile is proportional.
The function $a(\alpha)$ describes how the crack phase affects the initial elastic properties.
The function $w(\alpha)$ describes the local dissipation due to the crack phase growth.
The scaling constant $c_w$ is deriving from the choice of $w$.
For the choice of these functions, we retain the AT1 model, proposed by \textcite{pham_gradient_2011} and shown to be pertinent against experiments by \textcite{tanne_crack_2018}.
In the AT1 model, the functions are
\begin{equation}
  a(\alpha) = (1 - \alpha)^2
  , \quad
  w(\alpha) = \alpha
  , \quad
  c_w = 8/3.
\end{equation}

The energy functional  involved in the regularized minimization, in \cref{eq:phase field_minimization}, \emph{i.e.} the smeared crack model, has been shown to $\Gamma$-converge towards its sharp crack counterpart, in \cref{eq:griffith_principle}, when $\ell \rightarrow 0$ \parencite{braides_approximation_1998}.
In practice, it means that the solution of \cref{eq:phase field_minimization} converges to the solution of Griffith theory, in \cref{eq:griffith}, when $\ell \to 0$.
Therefore, the convergence should also be recovered on the numerical solution when the numerical errors vanish.

\section{Quasi-static equilibrium path}
\label{sec:path-following}

At this point, we presented two models for Griffith theory: the sharp crack model and the smeared crack model using phase field.
As specified in the introduction, this work focuses on the study of crack propagation in quasi-static conditions.
To track the whole crack propagation, the crack must propagate steadily: it must verify the Griffith criterion ($G=G_c$) throughout the crack propagation.
The series of mechanical states verifying $G=G_c$ during the whole crack propagation is called the quasi-static equilibrium path.
Following the equilibrium path provides a mean to simulate the whole crack propagation without instability.
In this work, it enables capturing what happens at the beginning of crack propagation, where potential instabilities can occur.
In practice, following this path consists in adapting the load during the crack propagation to stay at the crack propagation threshold.
In this section, we briefly show how the quasi-static equilibrium path of the structure can be recovered, with both the sharp crack and the phase field models.
More complete description and discussion of path-following methods in the context of phase field models for fracture are proposed in previous work \parencite{loiseau_path-following_2025}.

\subsection{Sharp crack model}
\label{sub:equilibrium_path_lefm}

To follow the quasi-static equilibrium path with the sharp crack model, the idea consists in adapting the load to verify the Griffith criterion $G = G_c$ throughout the crack propagation.
We employ the same method as \textcite{triclot_toughening_2024} to obtain the displacement field at equilibrium for different crack length.
It is recalled in the following paragraph.
Note that the crack path must be known \emph{a priori} to apply this method.

Let us consider an elastic body $\Omega$ with an imposed displacement $\uimp = \lambda \uimpbar$ on its boundary $\partial \Omega_{\uu}$, where $\lambda$ is the load factor (or amplitude).
Linear elasticity ensures that the whole displacement field in the body is proportional to the load factor: $\uu(\xx) = \lambda \widebar{\uu}(\xx)$, where $\widebar{\uu}(\xx)$ is the displacement field induced by the unitary prescribed displacement $\uimpbar$.
The domain contains an initial crack of length $a(t=0) = a_0$, which evolves during the (proportional) loading.
The crack propagates when the energy release rate $G$ reaches the critical energy release rate $G_c$.
As the energy release rate $G$ is quadratic in the displacement $\uu$, we can calculate it from the energy release rate $\widebar{G}$ corresponding to $\lambda = 1$ writing $G = \lambda^2 \widebar{G}$.
Therefore, the critical load factor $\lambda$ can be recovered for each crack length $a$ by:
\begin{enumerate}
  \item Solving the elastic problem to obtain the displacement field $\widebar{\uu}$ for a unitary load ($\lambda = 1$),
  \item Calculating the associated energy release rate $\widebar{G}(a)$,
  \item Calculating the critical load factor using Griffith criterion
    \begin{equation}
      \lambda^2(a) \widebar{G}(a) = G_c \implies \lambda(a) = \sqrt{\widebar{G}(a) / G_c}.
    \end{equation}
\end{enumerate}
Knowing the critical loads for any crack length $a$ enables to recover the quasi-static equilibrium path $\uu(\xx, a) = \lambda (a) \widebar{\uu}(\xx, a)$.

In practice, we employ Finite Element simulations to compute the displacement field $\widebar{\uu}(\xx, a)$ for a unitary load ($\lambda=1$) and a given crack length $a$.
The energy release rate $\widebar{G}$ is calculated using the $G(\Theta)$ method, originally proposed by \textcite{destuynder_sur_1981} and later discussed by \textcite{moran_crack_1987,suo_application_1992}.
The \cref{alg:equilibrium_path_lefm} summarizes the methods and its numerical implementation.

\begin{Algorithm}
\caption{Computation of the quasi-static equilibrium path for the sharp crack model using Finite Element elastic simulations.}
\label{alg:equilibrium_path_lefm}
\begin{algorithmic}[1]
  \Require Mesh containing a crack of length $a$, initial crack length $a_0$, crack length increment $\Delta a$ and the maximum crack length $a_{\mathrm{max}}$.
  \State Initialize the crack length $a = a_0$
  \While{$a < a_{\mathrm{max}}$}
    \State Solve the elastic problem for a unitary loading ($\lambda = 1$) $\to \widebar{\uu}(\xx)$
    \State Compute the (unitary) energy release rate $\to \widebar{G}$
    \State Compute the critical load factor $\to \lambda = \sqrt{\widebar{G} / G_c}$
    \State Compute the displacement field at equilibrium $\to \uu(\xx) = \lambda \widebar{\uu}(\xx)$
    \State Increment the crack length $\to a = a + \Delta a$
  \EndWhile
\end{algorithmic}
\end{Algorithm}

\subsection{Smeared crack model}

Due to the regularization of the crack discontinuity in phase field models, the method proposed in the previous section can not be directly applied.
To obtain the quasi-static equilibrium path in phase field models, the idea consists in applying an indirect load control method \parencite{rastiello_path-following_2022}.
Those methods, initially employed for geometric non-linearities \parencite{wempner_discrete_1971,riks_application_1972, riks_incremental_1979, crisfield_fast_1981}, consist in introducing the load factor $\lambda$ as an unknown in the minimization problem, and its associated control equation.
For crack propagation problem, the control equation is arbitrarily chosen to limit crack propagation and enforces a steady propagation.
For instance, \textcite{gutierrez_energy_2004} limited the elastic energy release between two load steps, whereas \textcite{singh_fracture-controlled_2016} directly controlled the fracture dissipation (which is equivalent to controlling the crack length).
Various control equations have been reviewed by \textcite{rastiello_path-following_2022}.
In this work, we chose to control the load factor $\lambda$ by imposing the maximum strain increment (over the domain $\Omega$) between two steps as proposed by \textcite{chen_secant_1991}.
It restricts the maximum strain increment in the domain between two load steps to limit the crack growth.
Given the state $(\uu_0, \alpha_0)$ at the previous load step, the minimization problem of \cref{eq:phase field_minimization} becomes
\begin{equation}
  \label{eq:phase field_with_path-following}
  \begin{split}
    (\widebar{\uu}, \alpha) = & \arg\min_{\substack{\uu \in \mathcal{U} \\ \alpha \in \mathcal{A}}}
      \mathcal{E}(\lambda \widebar{\uu}, \alpha) + \mathcal{D}(\alpha) - \mathcal{W}_{\mathrm{ext}} (\lambda \widebar{\uu}) \\
    & \text{subject to } \max_{\xx \in \Omega} (\eps(\lambda \widebar{u}(\xx)) - \eps(\uu_0)) = \Delta \eps_{\mathrm{imp}},
  \end{split}
\end{equation}
where $\Delta \eps_{\mathrm{imp}}$ is an arbitrary maximum strain increment which (indirectly) controls the load step.

To solve the problem in \cref{eq:phase field_with_path-following}, we extend the classic solution method of phase field fracture models.
Let us start by describing how the problem without path-following in \cref{eq:phase field_minimization} is solved.
As the functional is poly-convex, the minimization problem can be split into two sub-problems on the displacement field and the crack phase \parencite{bourdin_numerical_2000}.
The solution method consists of alternatively solving those two sub-problems, hence the alternate minimization name.
The equation governing the stationary points of each sub-problem is obtained by calculating the directional derivatives of the energy functional and equating them to zero for any direction.
It gives the variational formulation for the two sub-problems.
From a practical point of view, the discretization of the problem and its resolution are performed via the Finite Element Method using FEniCSx \parencite{alnaes_unified_2014,baratta_dolfinx_2023}.
The irreversibility constraint on the phase field is imposed through variational inequality using the solver \texttt{vinewtonrsls} from the PETSc library.

When introducing the path-following constraint in \cref{eq:phase field_with_path-following}, the problem can still be solved alternatively.
Instead of solving the displacement sub-problem for given boundary conditions, it is solved with unitary boundary conditions ($\lambda = 1 \implies \uimp = \uimpbar$) giving the displacement field $\widebar{\uu}(\xx)$.
Then, the control equation is solved to get the critical load factor $\lambda$, and the displacement field is rescaled: $\uu(\xx) = \lambda \widebar{\uu}(\xx)$.
The phase field sub-problem remains unchanged.
The algorithm to solve one load step of this problem is summarized in \cref{alg:equilibrium_path_pf}.

\begin{Algorithm}
\caption{Computation of a load step of the quasi-static equilibrium path for the smeared crack model.}
\label{alg:equilibrium_path_pf}
\begin{algorithmic}[1]
  \Require Domain mesh, initial state $(u_0(\xx), \alpha_0(\xx))$, load increment $\Delta \eps_{\mathrm{imp}}$.
  \While{not converged}
    \State Solve the elastic sub-problem for a unitary loading ($\lambda = 1$) $\to \widebar{\uu}(\xx)$
    \State Solve the control equation to obtain the critical load factor $\to \lambda$
    \State Compute the displacement field at equilibrium: $\to \uu(\xx) = \lambda \widebar{\uu}(\xx)$
    \State Solve the phase field sub-problem to the phase field $\to \alpha(\xx)$ \label{lst:phase_field_subproblem}
  \EndWhile
\end{algorithmic}
\end{Algorithm}

\section{Initial crack implementation techniques in phase field fracture}
\label{sec:crack_initialization_techniques}

This section presents the different crack initialization techniques for phase field fracture simulations that will be compared and analyzed.
The techniques are classified into two categories: the geometric initial cracks (explicitly represented in the mesh) and the phase initial cracks (embedded in the initial phase field).
For each category, different variants of the technique will be proposed.
An identifier is associated with each technique variant for clarity purposes.
\Cref{fig:crack_initialization_techniques} provides a summary and illustrates each of the variants considered in this work.

\begin{figure}
  \centering
  \input{./figure0.tex}
  \caption{Illustration of the different crack initialization techniques. The green line on the GEO-T0 illustrations corresponds to the line where the node are duplicated. The color corresponds to the phase field with blue being uncracked ($\alpha=0$) and red being fully cracked ($\alpha=1$). The abbreviation BC corresponds to boundary condition.}
  \label{fig:crack_initialization_techniques}
\end{figure}

\subsection{Geometric initial crack (GEO)}

The geometric initial cracks are specified in the geometry (\emph{i.e.}, in the mesh) of the domain.
This technique corresponds to incorporating initial cracks as sharp (non-regularized) discontinuities with traction-free boundary conditions, as it is usually done in fracture mechanics.

Different variants of the geometric technique will be tested, starting with changes of the crack thickness.
Two thicknesses are compared: an infinitely thin crack (T0) and a one-element-wide crack (T1).
Note that infinitely thin crack (T0) corresponds to duplicating nodes along the crack path, as often employed in fracture mechanics Finite Element simulations.
Then, the effect of boundary conditions on the initial crack is also investigated.
Three different boundary conditions in crack phase $\alpha$ are considered: zero Neumann boundary conditions (NEU), crack phase imposed to 1 at the crack tip (TIP), and crack-phase imposed to 1 along whole crack (WHL).

\subsection{Phase field initial cracks (PHA)}

The phase field initial cracks are obtained by initializing the crack phase field.
It corresponds to adding the initial cracks to the domain as regularized discontinuities.

In this technique, the width of initial crack will also be investigated for the phase field initial crack.
Two variants are considered: the case of the infinitely thin initial crack (T0), and the case of a one-element-wide initial crack (T1).

Let us describe how the phase field initial crack are implemented in practice.
For clarity purposes, we consider the case of a single straight initial crack.
Curved initial cracks are discussed in \cref{sec:discussion}.
To implement phase field initial cracks, two steps are required.
First, the nodal values of the phase field are set to one along the initial crack.
The case of an infinitely thin (T0) and one-element-wide (T1) initial crack must be distinguished.
For the (T0) initial crack, the phase field $\alpha(\xx) = 1$ is prescribed at the nodes along the initial crack.
It requires the mesh nodes to be rigorously aligned along the initial crack.
For a (T1) initial crack, the phase field $\alpha(\xx) = 1$ is prescribed on the nodes of elements crossed by the initial crack.
In this case, the elements must be aligned to represent a straight crack accurately so that the initial crack goes through their centers.
The second step consists in regularizing the sharp crack to make it smeared.
This can be done by solving the phase field sub-problem (see step \ref{lst:phase_field_subproblem} of \cref{alg:equilibrium_path_pf}) with zero loads.

\begin{figure}
  \centering
  \begin{subfigure}{0.49\textwidth}
    \centering
    \includegraphics[width=\textwidth]{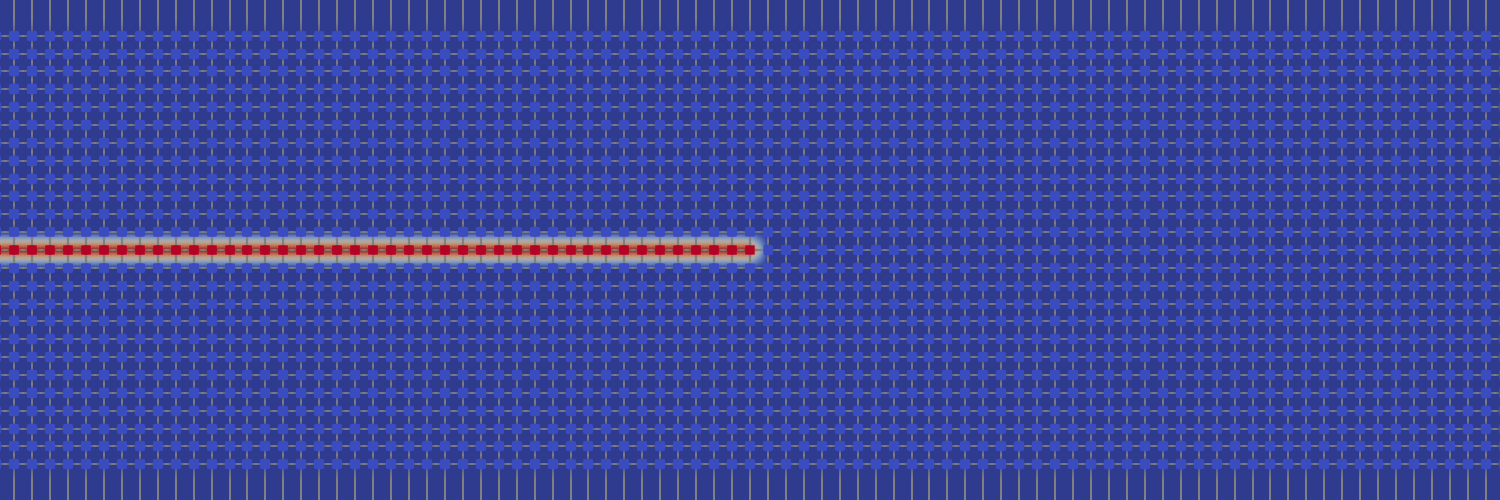}
    \caption{Infinitely thin (T0), before regularization.}
  \end{subfigure}
  \begin{subfigure}{0.49\textwidth}
    \centering
    \includegraphics[width=\textwidth]{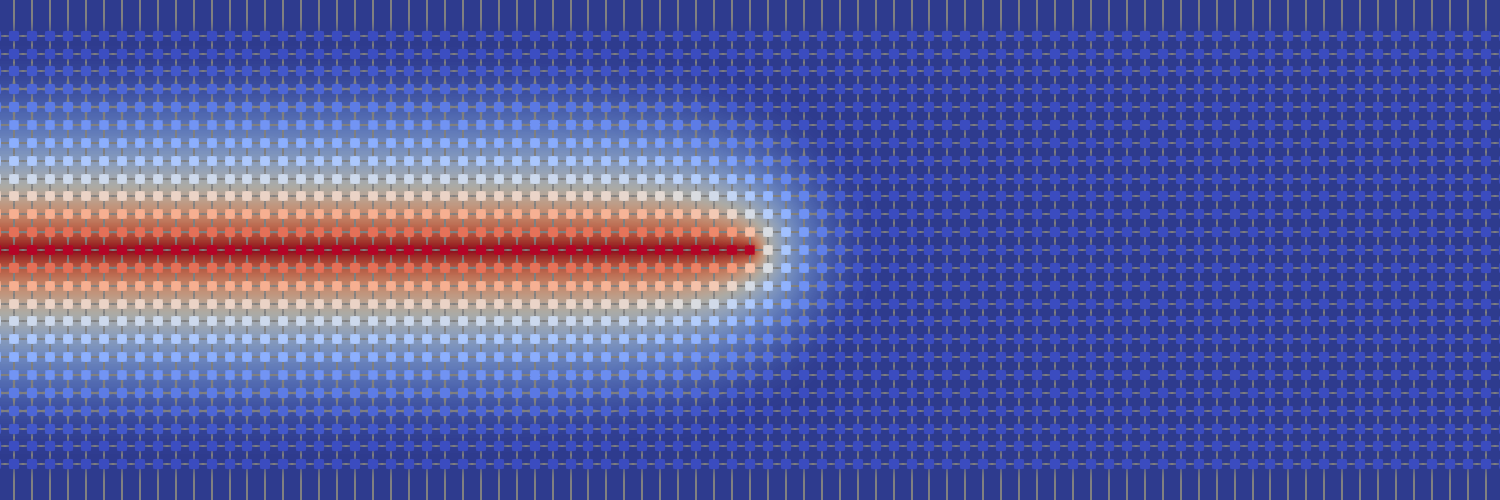}
    \caption{Infinitely thin (T0), after regularization.}
    \label{fig:definition_initial_cracks_PHA_T0_post_reg}
  \end{subfigure}
  \begin{subfigure}{0.49\textwidth}
    \centering
    \includegraphics[width=\textwidth]{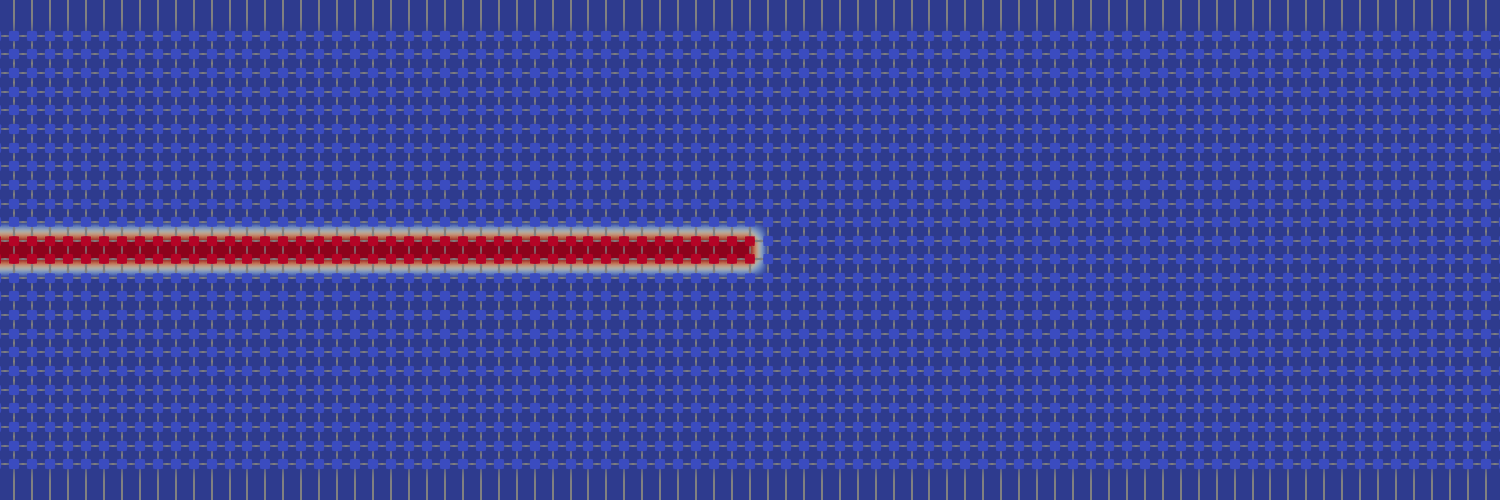}
    \caption{One-element-wide (T1), before regularization.}
  \end{subfigure}
  \begin{subfigure}{0.49\textwidth}
    \centering
    \includegraphics[width=\textwidth]{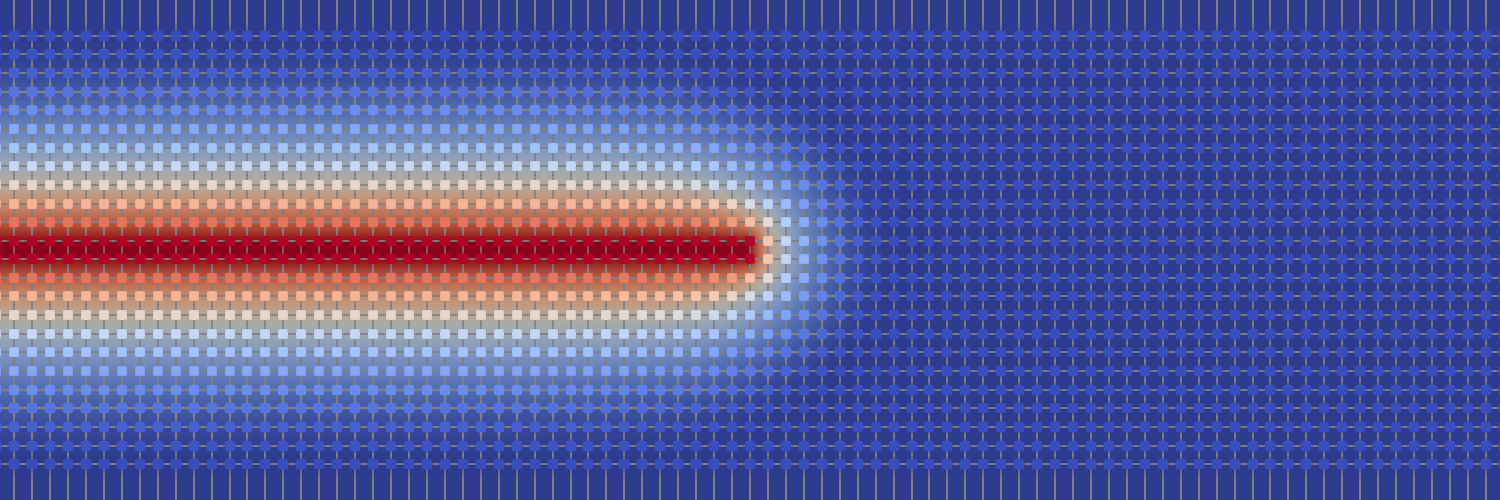}
    \caption{One-element-wide (T1), after regularization.}
  \end{subfigure}
  \caption{Illustration of the definition of phase field initial crack (PHA). The red color corresponds to the crack ($\alpha$) whereas the blue color corresponds to the uncracked parts ($\alpha=0$).}
  \label{fig:definition_initial_cracks_PHA}
\end{figure}

\Cref{fig:definition_initial_cracks_PHA} illustrates how the initial phase field is defined for both variants.
This figure shows the phase field for infinitely thin and one-element-wide initial cracks.
The values before and after regularization are shown.
The nodal values of the phase field are represented as square markers in the figure to highlight the prescribed field values before the regularization.

\section{Comparison of initial crack implementation techniques}
\label{sec:comparison_crack_initialization_techniques}

\subsection{Presentation of the numerical benchmark}

\begin{figure}
  \centering
  \begin{tikzpicture}[scale=5]
  \def\L{1} 
  \def\a{0.5} 
  
  \draw[thick] (0,0) rectangle (\L,\L);

  \draw[ultra thick] (0,\L/2) -- (\a,\L/2);
  
  \draw[thick, latex-latex] (-0.1,0) -- (-0.1,\L) node[midway, left] {$L = 1$\,mm};
  \draw[thick, latex-latex] (0,\L/2+0.05) -- (\a, \L/2+0.05) node[midway, above] {$a_0 = 0.5$\,mm};

  \foreach \x in {0.0, 0.2,..., \L} {
    \draw[thick, -latex] (\x,\L+0.01) -- (\x,\L+0.16);
  }
  \draw[thick] (0,\L+0.16) -- (\L,\L+0.16) node[midway, above] {$\uimp$};

  \draw[thick] (0,0) -- ({-0.1*sin(30)},{-0.1*cos(30)}) -- ({+0.1*sin(30)},{-0.1*cos(30)}) -- (0,0);
  \filldraw[thick, fill=white, draw=black] (0,0) circle (0.5pt);
  \draw[thick] ({-0.1*sin(30)+0.025},{-0.1*cos(30)}) -- ({-0.1*sin(30)-0.005},{-0.1*cos(30)-0.03});
  \draw[thick] ({-0.1*sin(30)+0.055},{-0.1*cos(30)}) -- ({-0.1*sin(30)+0.025},{-0.1*cos(30)-0.03});
  \draw[thick] ({-0.1*sin(30)+0.085},{-0.1*cos(30)}) -- ({-0.1*sin(30)+0.055},{-0.1*cos(30)-0.03});

  \foreach \dx in {0.2, 0.4,...,\L} {
    \draw[thick] (\dx,0) -- ({-0.1*sin(30)+\dx},{-0.1*cos(30)}) -- ({+0.1*sin(30)+\dx},{-0.1*cos(30)}) -- (\dx,0);
    \filldraw[thick, fill=white, draw=black] (\dx,0) circle (0.5pt);
    \filldraw[thick, fill=white, draw=black] ({\dx-0.045},{-0.1*cos(30)-0.02}) circle (0.5pt);
    \filldraw[thick, fill=white, draw=black] (\dx,{-0.1*cos(30)-0.02}) circle (0.5pt);
    \filldraw[thick, fill=white, draw=black] ({\dx+0.045},{-0.1*cos(30)-0.02}) circle (0.5pt);
  }
  \draw[thick] (0.1,{-0.1*cos(30)-0.043}) -- (1.1,{-0.1*cos(30)-0.043});
\end{tikzpicture}
  \caption{Geometry of the SENT specimen and boundary conditions.}
  \label{fig:sent_specimen}
\end{figure}
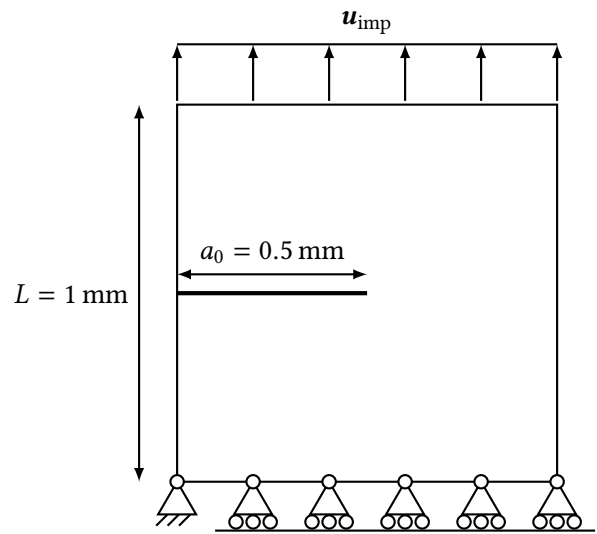

The Single Edge Notched Tensile (SENT) test, which is often used a numerical benchmark for phase field simulations,  is employed to compare the different crack initialization techniques.
The geometry and boundary conditions, illustrated in \cref{fig:sent_specimen}, correspond to those used by \textcite{singh_fracture-controlled_2016}.
The SENT specimen is a square with a side length $L = \qty{1}{\milli\meter}$ and an initial crack of length $a_0 = L/2 = \qty{0.5}{\milli\meter}$.
The material is linear elastic with a Young modulus $E=\qty{230.77}{\giga\pascal}$, a Poisson ratio $\nu=0.43$, a critical energy release rate $G_c=\qty{2 700}{\joule\per\square\meter}$, and a regularization length $\ell=\qty{0.015}{\milli\meter}$.
In this section, we use two types of mesh for the phase field simulations.
First, an ideal structured mesh with elements aligned with the initial crack and the expected crack path is used to concentrate fully on the influence of the initial crack implementation technique.
The techniques satisfyingly representing initial cracks with this mesh are then confronted with an unstructured mesh (still verifying the conditions required to apply the technique).
For the sake of brevity, the meshes are represented in the results.
The mesh size is set to $h=\ell/6$ in the region $y \in [L/2 - 2 \ell, L/2 + 2 \ell]$, where $y$ is the coordinate along the vertical direction.
Larger elements with a size $h_{\mathrm{far}} = 2 \ell$ are used outside this region.

Following the recommendations of \textcite{bourdin_variational_2008}, the critical energy release rate in the phase field simulations is adjusted to obtain the correct effective energy release rate 
\begin{equation}
    G_c^{\mathrm{eff}} = G_c^{\mathrm{PF}} \left( 1 + \frac{h}{c_w \ell} \right).
\end{equation}
Hence, in the phase field simulations, we choose the critical energy release rate $G_c^{\mathrm{PF}}$ so that the effective critical energy release rate $G_c^{\mathrm{eff}} = G_c \implies G_c^{\mathrm{PF}} = \qty{2541}{\joule\per\square\meter}$.

To compare the techniques, we will consider the force-displacement curves from the sharp and smeared crack models in \cref{fig:Fu_T0,fig:Fu_T1,fig:Fu_unstructured}.
In these curves, the displacement $u$ is the vertical component of the imposed displacement $\uimp$.
The force $F$ is the vertical component of the reaction force $\bm{F}$ calculated by integrating the traction vector (obtained by post-processing the displacement field $\uu$) over the upper face.

\subsection{Infinitely thin initial cracks (T0)}

\begin{figure}
  \centering
  \begin{subfigure}{\textwidth}
    \centering
    \input{./figure3.tex}
    \caption{Force-displacement curves for an infinitely thin initial crack (T0). The inset illustrates thickening of the crack phase band occurring for PHA-T0 during the snap-back.}
    \label{fig:Fu_T0}
  \end{subfigure}
  \begin{subfigure}{0.49\textwidth}
    \centering
    \includegraphics[width=\textwidth]{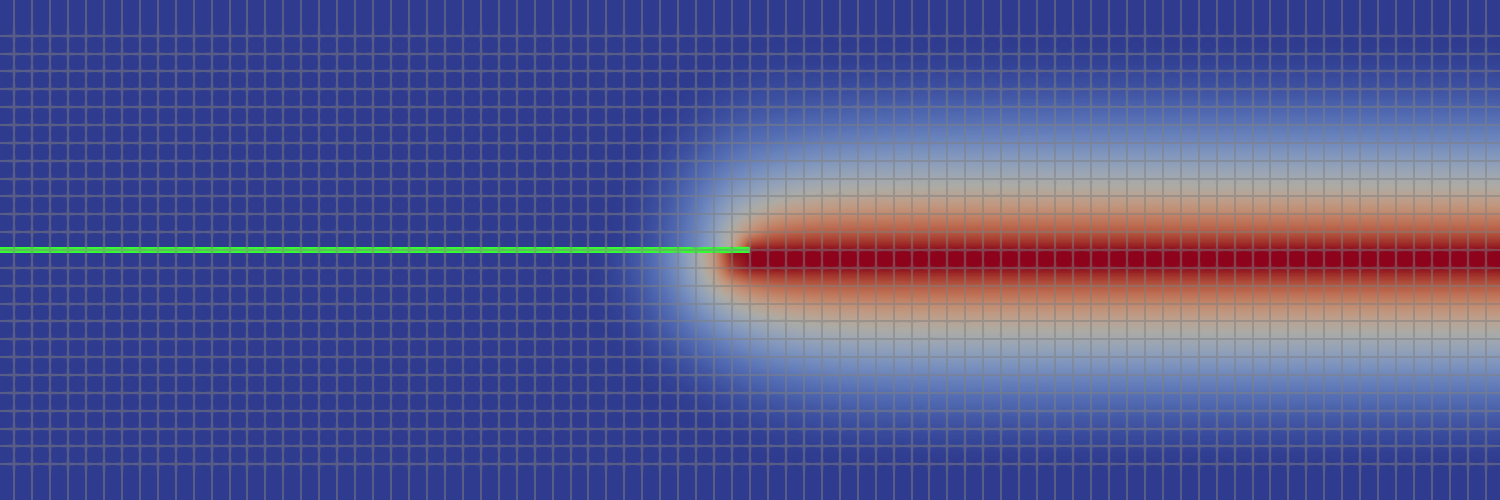}
    \caption{GEO-T0-NEU}
    \label{fig:crack_field_GEO-T0-NEU_tf}
  \end{subfigure}
  \begin{subfigure}{0.49\textwidth}
    \centering
    \includegraphics[width=\textwidth]{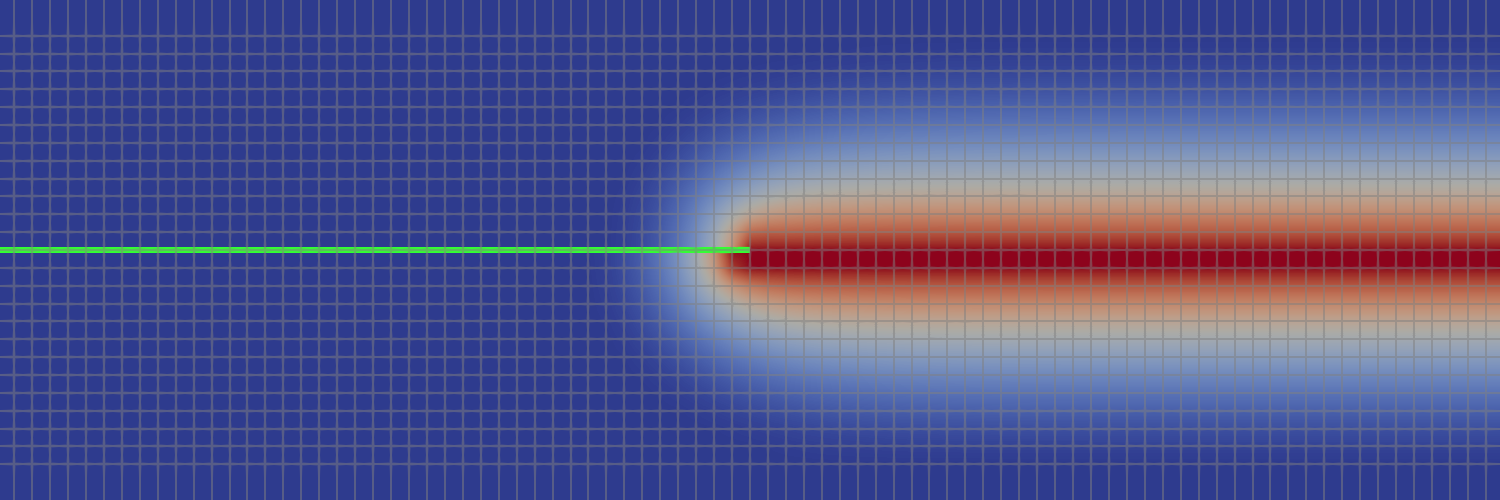}
    \caption{GEO-T0-TIP}
    \label{fig:crack_field_GEO-T0-TIP_tf}
  \end{subfigure}
  \begin{subfigure}{0.49\textwidth}
    \centering
    \includegraphics[width=\textwidth]{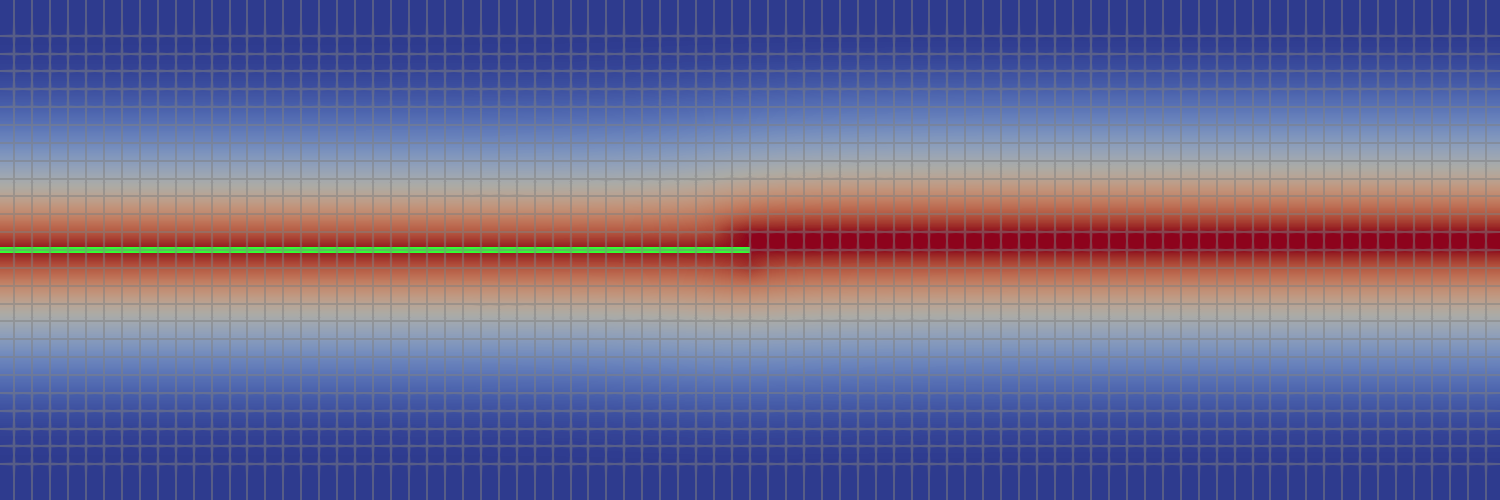}
    \caption{GEO-T0-WHL}
    \label{fig:crack_field_GEO-T0-WHL_tf}
  \end{subfigure}
  \begin{subfigure}{0.49\textwidth}
    \centering
    \includegraphics[width=\textwidth]{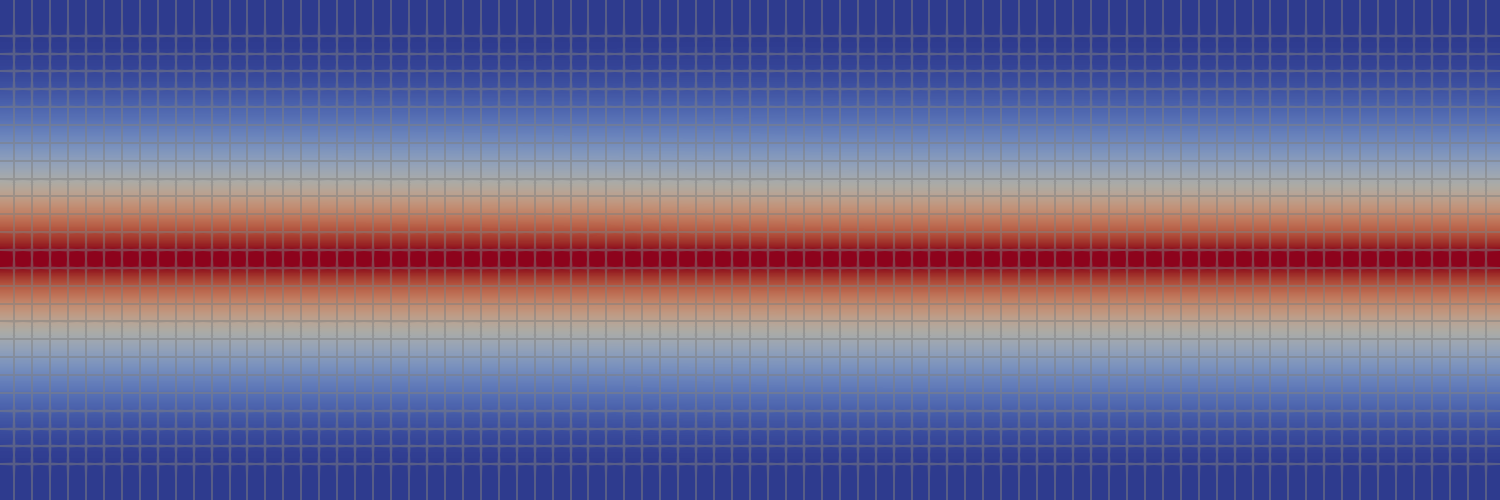}
    \caption{PHA-T0}
    \label{fig:crack_field_PHA-T0_tf}
  \end{subfigure}
  \caption{Comparison of the phase field fracture solutions with the different crack initialization techniques for an infinitely thin initial crack (T0). The top part shows the force-displacement curves along with the reference solution from the sharp crack model. The lower part shows the crack phase field after crack propagation with blue being uncracked ($\alpha=0$) and red being fully cracked ($\alpha=1$).}
  \label{fig:solutions_T0}
\end{figure}

We start by presenting the results with infinitely thin initial cracks (variants with T0).
The comparison will be based on two types of observations both represented on \cref{fig:solutions_T0}.
The first type of observations, in \cref{fig:Fu_T0}, is the force-displacement curves.
The force-displacement obtained with the different initial crack implementation techniques
are represented along with the reference curve of the sharp crack model.
The second type of observation is the phase field around the initial crack tip after the propagation.
They are provided in \cref{fig:crack_field_GEO-T0-NEU_tf,fig:crack_field_GEO-T0-TIP_tf,fig:crack_field_GEO-T0-WHL_tf,fig:crack_field_PHA-T0_tf}.

Before commenting the peak and post-peak part of \cref{fig:Fu_T0}, let us note that the elastic phase is biased by the smeared crack model.
Indeed, the diffusion of the crack induces a loss of macroscopic rigidity.
This loss decreases when reducing the regularization length $\ell$.
This applies for all the following results (except for the PHA-T0 technique, which is discussed later).

The first observation, on \cref{fig:Fu_T0}, is that all the geometric methods (GEO) present an invalid peak on the force-displacement curve.
The peak reduces when the crack phase is set to 1 either on the crack tip or on the whole crack boundary.
Nevertheless, it persists with all the geometric methods, even when the crack phase is imposed to one on the whole crack (WHL).
In the \cref{fig:crack_field_GEO-T0-NEU_tf,fig:crack_field_GEO-T0-TIP_tf,fig:crack_field_GEO-T0-WHL_tf}, it can be observed that the crack phase is initially equal to one on a (duplicated) line of nodes.
However, after the propagation, the crack phase equals one on a line of elements (\emph{i.e.}, two lines of nodes).
Indeed, for the displacement ``discontinuity'' to occur in an element, the whole element needs to be cracked.
Thus, a slight bifurcation of the crack is necessary to transition from an infinitely thin initial crack to a one-element-wide crack that enables the displacement jump.
This transition requires an artificial excess of energy to propagate the infinitely thin initial crack compared to a proper initial crack.

The case of the PHA initial crack is different.
The associated force-displacement curve in \cref{fig:Fu_T0} shows a Young modulus that is initially too high and a snap-back is observed in what is supposed to be the elastic phase.
As shown in the inset images of \cref{fig:Fu_T0}, the observed snap-back corresponds to the thickening of the initial crack.
The reason is that a whole line of elements must be cracked for the displacement ``discontinuity'' to occur.
Thus, the initial crack field needs to thicken to allow the displacement jump across the crack.
Once this thickening is achieved, the remainder of the curves matches the reference results, and no force overshoot is observed.

From those first results, we can conclude that infinitely thin initial cracks should not be used in phase field fracture.
It means that a duplication of the nodes on the crack faces is not sufficient to adequately represent an initial crack (even when setting the crack phase to one along the crack).
Note that the proposition of \textcite{tanne_crack_2018} to introduce the initial crack as a sharp V-notch will show similar issues if the notch tip is one node.
Indeed, the problem of having to transition from  one node to a line of cracked elements persists.

\subsection{One-element-wide initial cracks (T1)}

\begin{figure}
  \centering
  \begin{subfigure}{\textwidth}
    \centering
    \begin{tikzpicture}
    \begin{axis}[
        width=0.9\textwidth,
        xlabel={Displacement $u$ [m]},
        ylabel={Force $F$ [N]},
        xmin=0, ymin=0,
        grid=both,
        legend pos=north west,
        legend cell align={left},
        set layers,
        mark size=3,
    ]
        \begin{pgfonlayer}{axis foreground}
        \addplot [very thick, no markers, color=semiana] table [x=u, y=F, col sep=comma] {./ref_lefm.csv};
        \end{pgfonlayer}
        \addlegendentry{Sharp crack model}
        \addplot [thick, color=tabblue, mark=x, mark repeat=100, mark phase=4] table [x=u, y=F, col sep=comma] {./GEO-T1-NEU.csv};
        \addlegendentry{GEO-T1-NEU}
        \addplot [thick, color=taborange, mark=+, mark repeat=100, mark phase=8] table [x=u, y=F, col sep=comma] {./GEO-T1-TIP.csv};
        \addlegendentry{GEO-T1-TIP}
        \addplot [thick, color=tabgreen, mark=square, mark repeat=100, mark phase=12] table [x=u, y=F, col sep=comma] {./GEO-T1-WHL.csv};
        \addlegendentry{GEO-T1-WHL}
        \addplot [thick, color=tabgray, mark=o, mark repeat=100, mark phase=44] table [x=u, y=F, col sep=comma] {./PHA-T1.csv};
        \addlegendentry{PHA-T1}
    \end{axis}
\end{tikzpicture}
    \caption{Force-displacement curves for a one-element-wide initial crack (T1).}
    \label{fig:Fu_T1}
  \end{subfigure}
  \begin{subfigure}{0.49\textwidth}
    \centering
    \includegraphics[width=\textwidth]{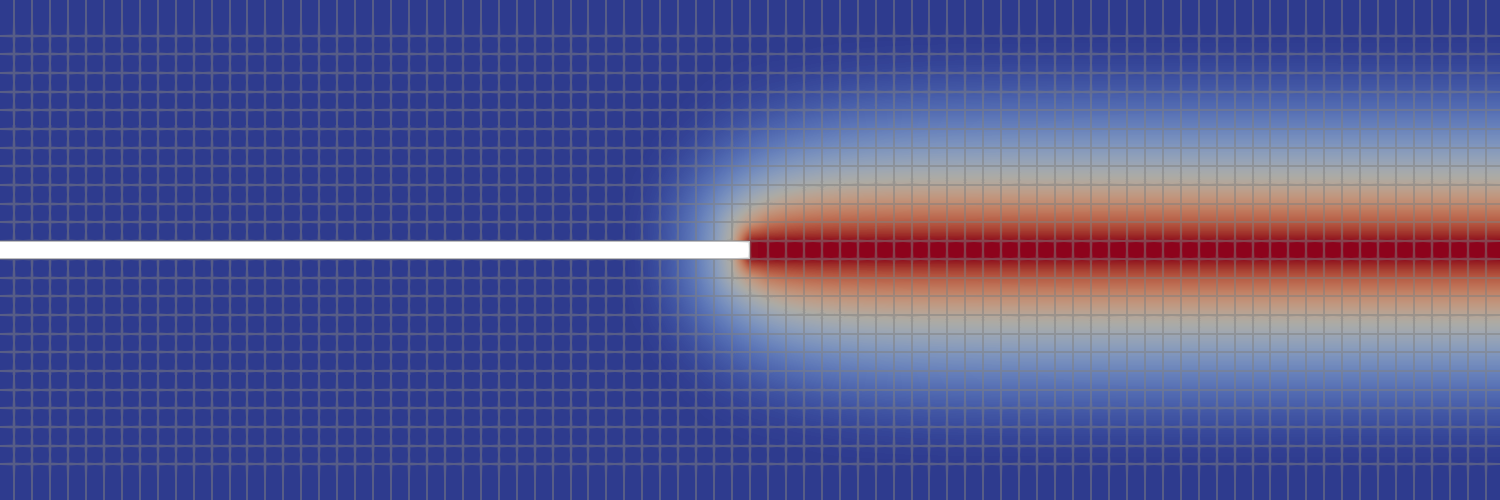}
    \caption{GEO-T1-NEU}
    \label{fig:crack_field_GEO-T1-NEU_tf}
  \end{subfigure}
  \begin{subfigure}{0.49\textwidth}
    \centering
    \includegraphics[width=\textwidth]{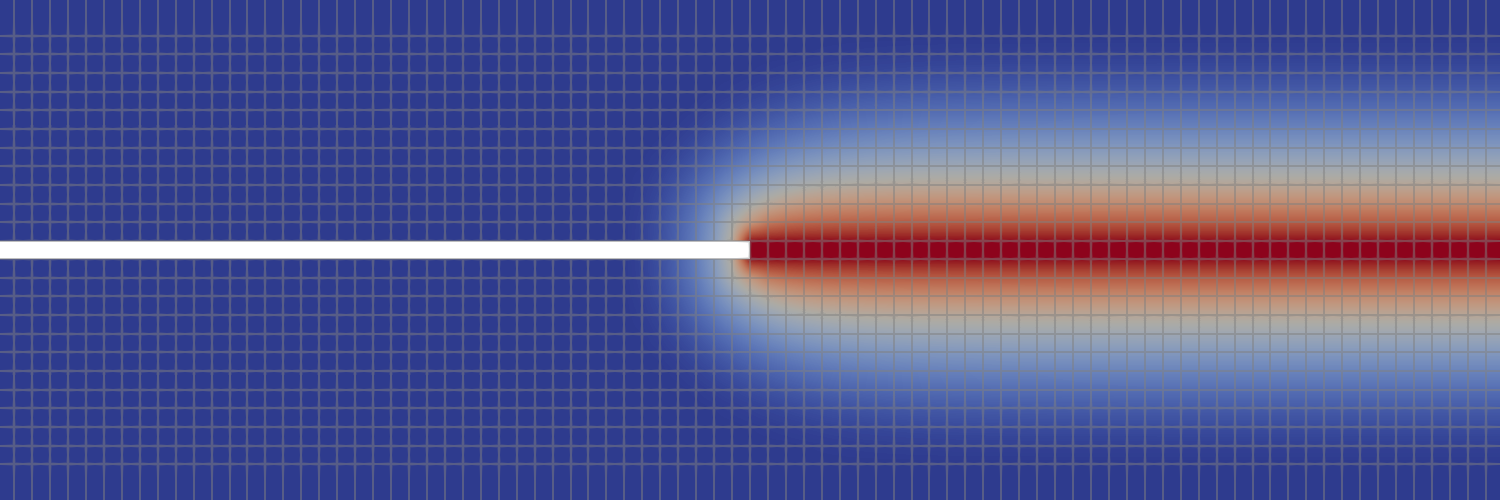}
    \caption{GEO-T1-TIP}
    \label{fig:crack_field_GEO-T1-TIP_tf}
  \end{subfigure}
  \begin{subfigure}{0.49\textwidth}
    \centering
    \includegraphics[width=\textwidth]{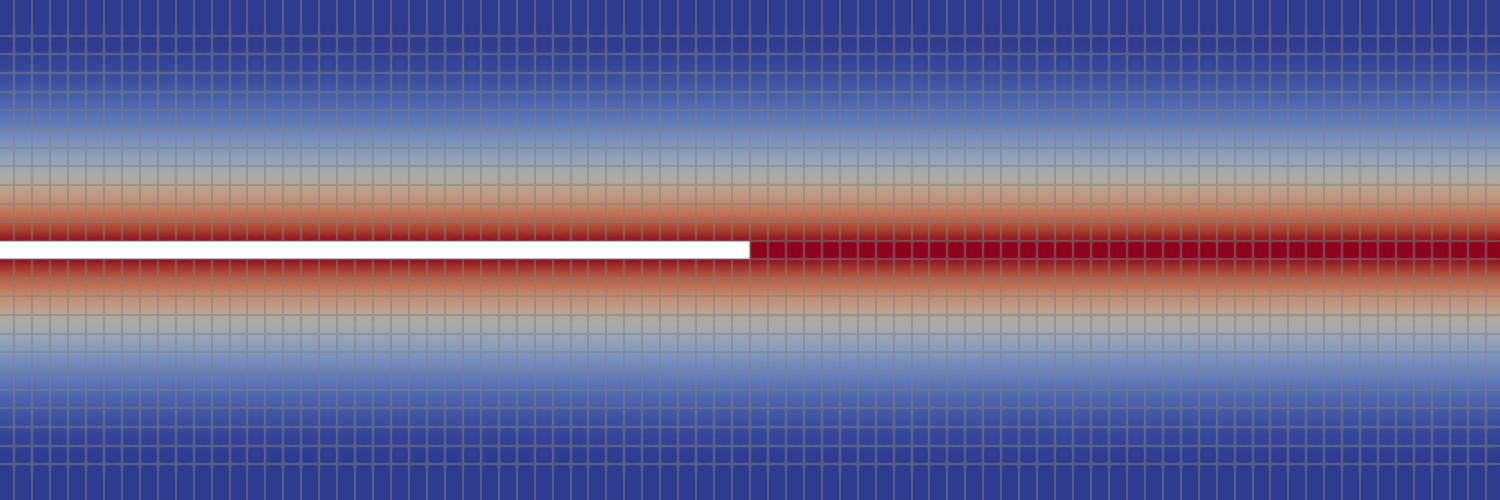}
    \caption{GEO-T1-WHL}
    \label{fig:crack_field_GEO-T1-WHL_tf}
  \end{subfigure}
  \begin{subfigure}{0.49\textwidth}
    \centering
    \includegraphics[width=\textwidth]{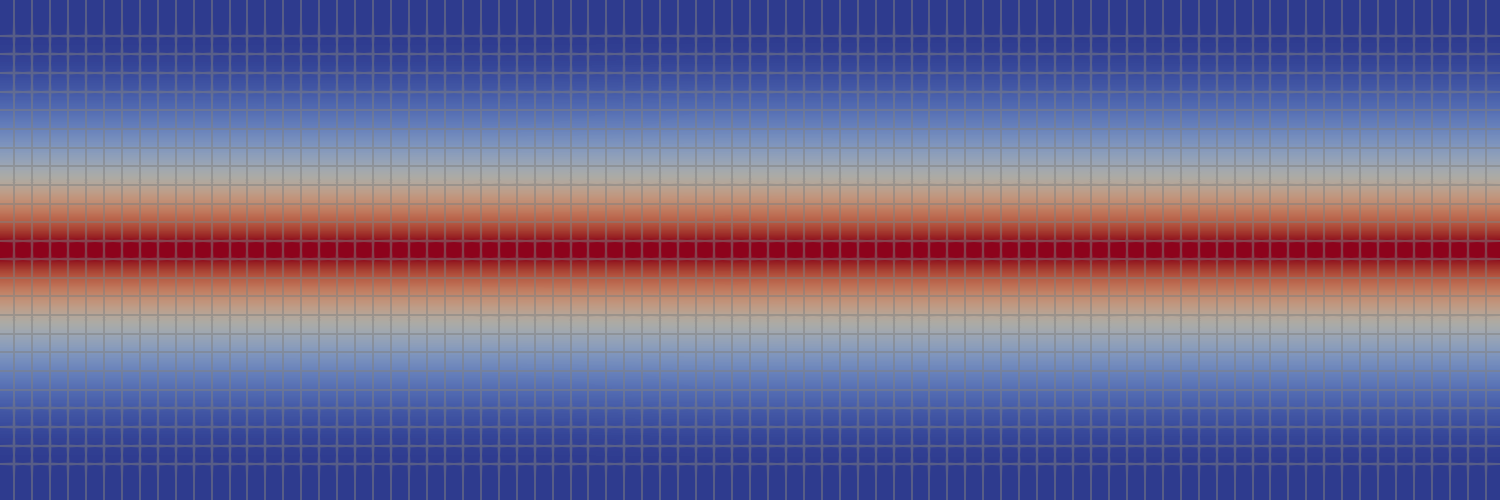}
    \caption{PHA-T1}
    \label{fig:crack_field_PHA-T1_tf}
  \end{subfigure}
  \caption{Comparison of the phase field fracture solutions with the different crack initialization techniques for a one-element-wide initial crack (T1). The top part shows the force-displacement curves along with the reference solution from the sharp crack model. The lower part shows the crack phase field after crack propagation with blue being uncracked ($\alpha=0$) and red being fully cracked ($\alpha=1$).}
  \label{fig:solutions_T1}
\end{figure}

Let us now present the results for one-element-wide initial cracks.
The results are shown in the \cref{fig:solutions_T1}.
The \cref{fig:Fu_T1} provides a quantitative comparison of the initial crack implementation techniques through the force-displacement curves.
Reference results obtained via the sharp crack model are also reported in this figure.
The phase fields around the initial crack tip after propagation are also provided in \cref{fig:crack_field_GEO-T1-NEU_tf,fig:crack_field_GEO-T1-TIP_tf,fig:crack_field_GEO-T1-WHL_tf,fig:crack_field_PHA-T1_tf}.

The technique GEO-T1-NEU shows a significant overshoot on the force-displacement curve.
This overshoot is due to the use of the Neumann boundary conditions.
Indeed, it imposes that the derivative of the phase field normal to the crack is null, forcing multiple elements to crack at once to start the crack propagation.
Once again, it generates an artificial increase in the energy required for the crack to propagate.

The GEO-T1-TIP also shows a (smaller) overshoot on the force-displacement curve.
On \cref{fig:crack_field_GEO-T1-TIP_tf}, we observe a cone shape at the beginning of the crack propagation.
It can be compared to other methods where the overshoot is not observed, \emph{e.g.} GEO-T1-WHL on \cref{fig:crack_field_GEO-T1-WHL_tf}.
One possible explanation is that the regularization region of the phase is not developed as it would for a propagated crack.
What happens before the crack tip likely affects the energy required to propagate the initial crack, hence the crack propagation threshold.

\begin{figure}
  \centering
  \begin{tikzpicture}
    \begin{axis}[
        width=\textwidth,
        height=0.5\textwidth,
        xlabel={Load step},
        ylabel={Number of iteration},
        xmin=0, ymin=0,
        grid=both,
        legend pos= north east,
        legend cell align={left},
        mark size=0.75,
        no markers,
    ]
        \addplot [const plot, color=taborange] table [x=load_step, y=N_iter, col sep=comma] {./iter_GEO-T1-WHL.csv};
        \addlegendentry{GEO-T1-WHL}
        \addplot [const plot, color=tabgray] table [x=load_step, y=N_iter, col sep=comma] {./iter_PHA-T1.csv};
        \addlegendentry{PHA-T1}
    \end{axis}
\end{tikzpicture}
  \caption{Number of iteration per load step for the geometrical and embedded methods with a 1-element-wide initial crack.}
  \label{fig:initial_crack_N_iter}
\end{figure}
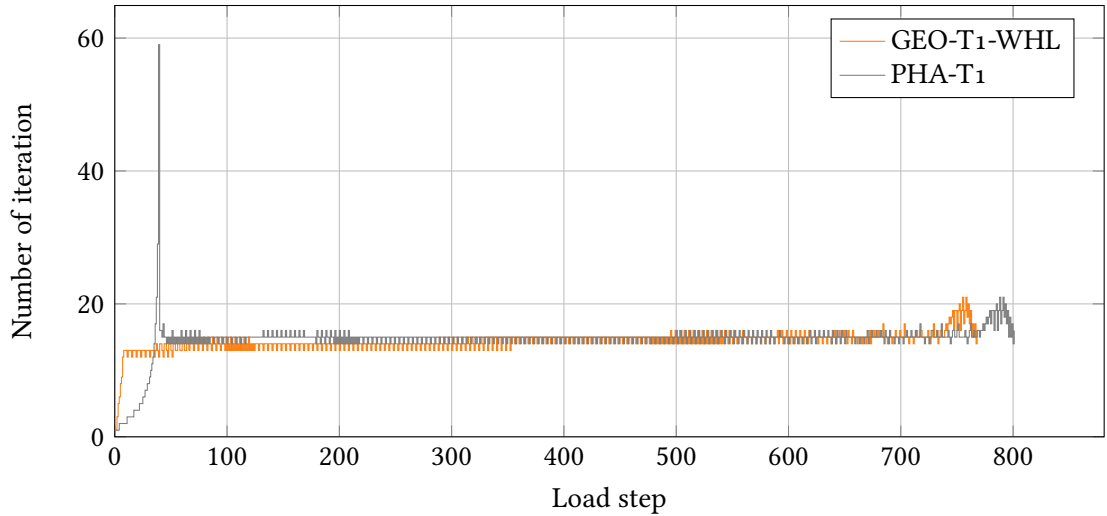

The crack initialization techniques GEO-T1-WHL and PHA-T1 can not be distinguished from each other in terms of mechanical response and phase field.
Their force-displacement curves have the same shape as the reference solution.
Moreover, we can observe, on \cref{fig:crack_field_GEO-T1-WHL_tf,fig:crack_field_PHA-T1_tf}, that the initial crack and the propagated part of the crack are indistinguishable for both implementation techniques.
To find which method is the best choice, we investigated the numerical performances of both methods.
The associated curve is represented on the \cref{fig:initial_crack_N_iter}.
The number of iterations required for the alternate minimization to converge gives an indication of the overall numerical performances of both methods.
\Cref{fig:initial_crack_N_iter} shows the alternate minimization number of iterations for each load step.
The geometric initialization technique cumulates 11 055 iterations, whereas the phase initialization technique has 11 734.
While the geometric method has a lower total number of iterations, no method clearly outperforms the other.

\subsection{Extension to unstructured mesh}
\begin{figure}
  \centering
  \begin{subfigure}{\textwidth}
    \centering
    \begin{tikzpicture}
    \begin{axis}[
        width=0.9\textwidth,
        xlabel={Displacement $u$ [m]},
        ylabel={Force $F$ [N]},
        xmin=0, ymin=0,
        grid=both,
        legend pos=north west,
        legend cell align={left},
        set layers,
        mark size=3,
    ]
        \begin{pgfonlayer}{axis foreground}
        \addplot [very thick, no markers, color=semiana] table [x=u, y=F, col sep=comma] {./ref_lefm.csv};
        \end{pgfonlayer}
        \addlegendentry{Sharp crack model}
        \addplot [thick, color=tabgreen, mark=square, mark repeat=100, mark phase=12] table [x=u, y=F, col sep=comma] {./GEO-T1-WHL_UNSTRUCTURED.csv};
        \addlegendentry{GEO-T1-WHL}
        \addplot [thick, color=tabgray, mark=o, mark repeat=100, mark phase=44] table [x=u, y=F, col sep=comma] {./PHA-T1_UNSTRUCTURED.csv};
        \addlegendentry{PHA-T1}
    \end{axis}
\end{tikzpicture}
    \caption{Force-displacement curves for a one-element-wide initial crack.}
    \label{fig:Fu_unstructured}
  \end{subfigure}
  \begin{subfigure}{0.49\textwidth}
    \centering
    \includegraphics[width=\textwidth]{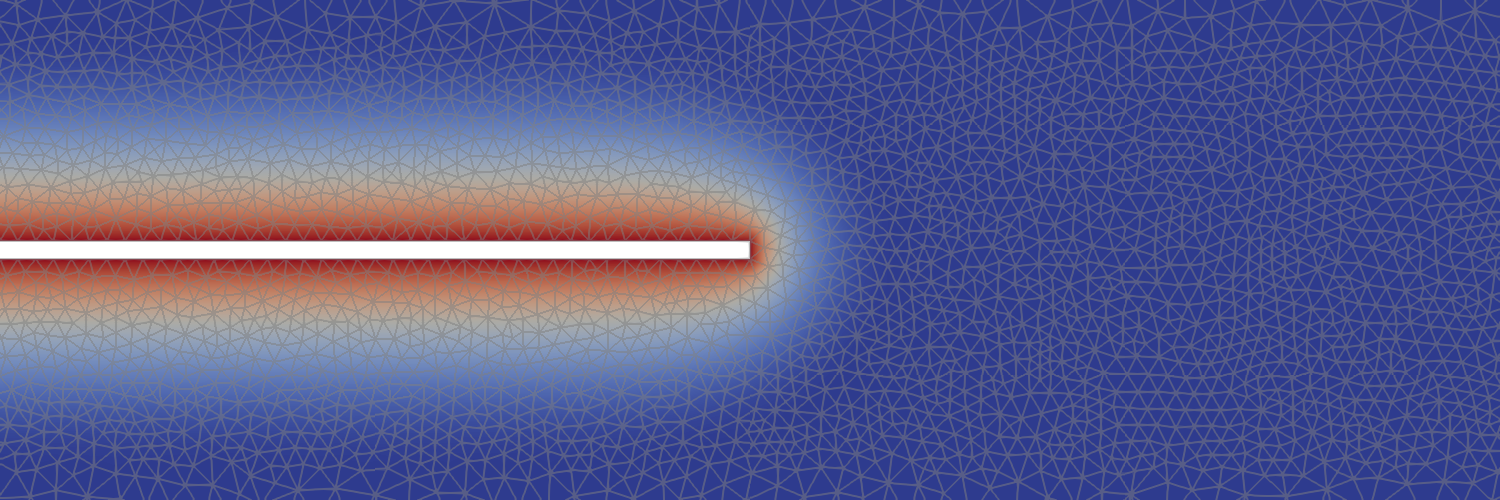}
    \caption{GEO-T1-WHL (before propagation)}
    \label{fig:crack_field_GEO-T1-WHL_UNSTRUCTURED_t0}
  \end{subfigure}
  \begin{subfigure}{0.49\textwidth}
    \centering
    \includegraphics[width=\textwidth]{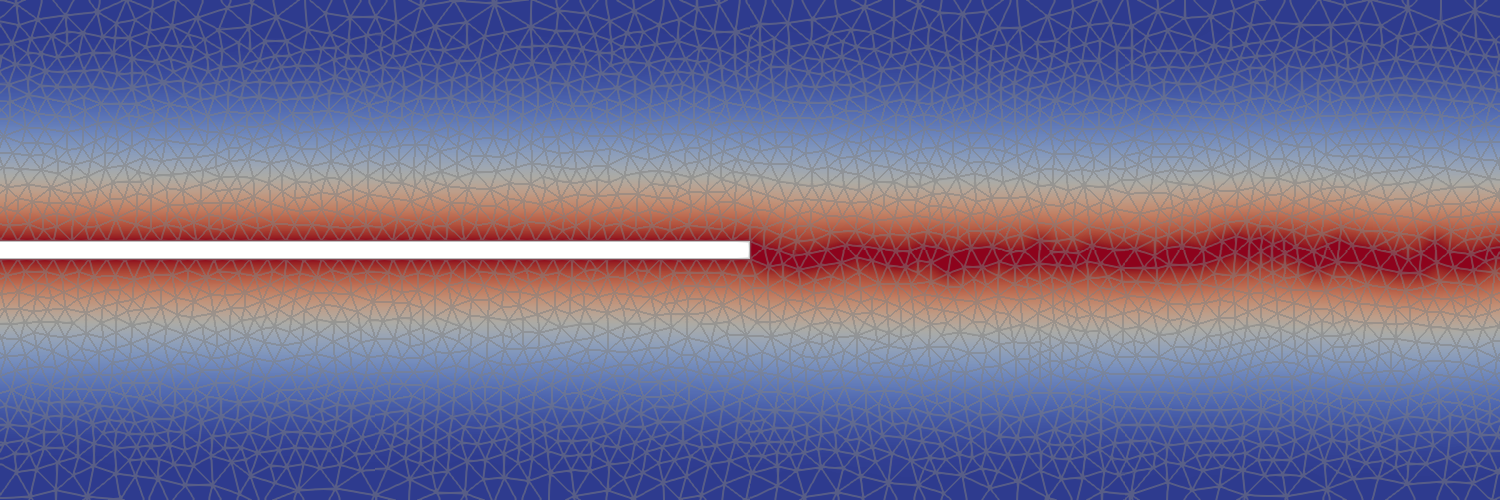}
    \caption{GEO-T1-WHL (after propagation)}
    \label{fig:crack_field_GEO-T1-WHL_UNSTRUCTURED_tf}
  \end{subfigure}
  \begin{subfigure}{0.49\textwidth}
    \centering
    \includegraphics[width=\textwidth]{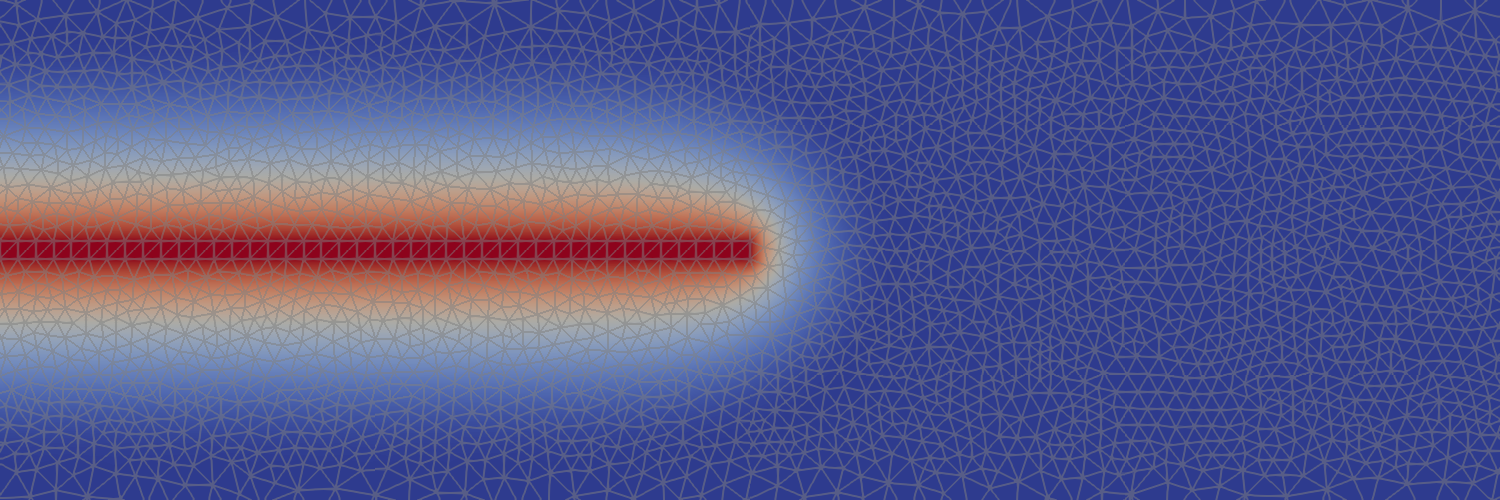}
    \caption{PHA-T1 (before propagation)}
    \label{fig:crack_field_PHA-T1_UNSTRUCTURED_t0}
  \end{subfigure}
  \begin{subfigure}{0.49\textwidth}
    \centering
    \includegraphics[width=\textwidth]{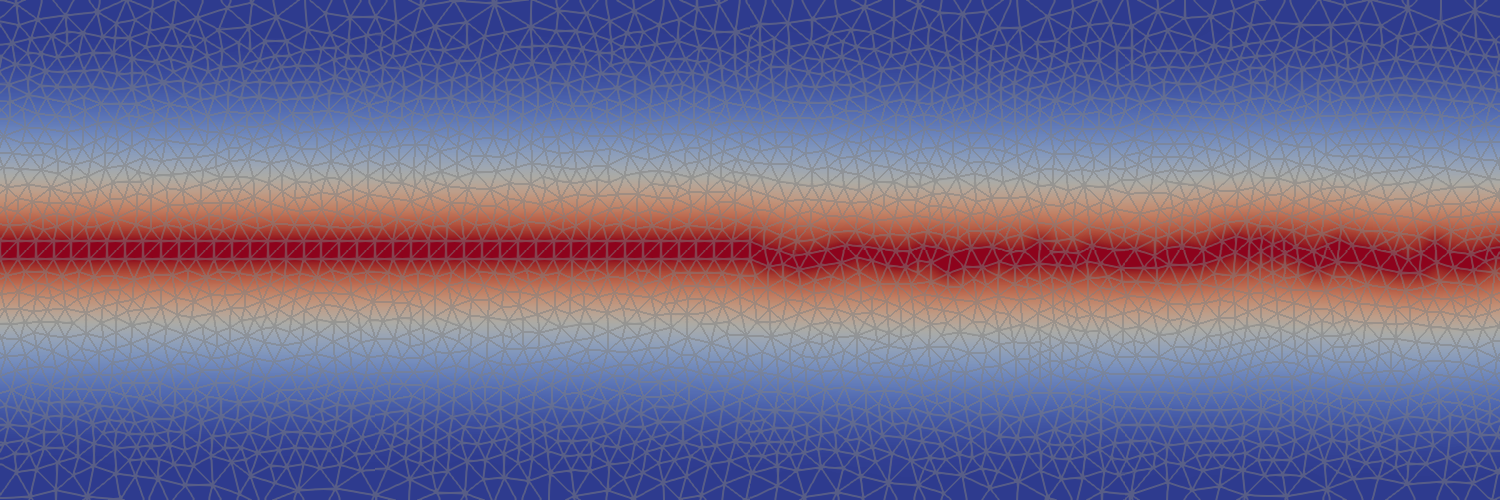}
    \caption{PHA-T1 (after propagation)}
    \label{fig:crack_field_PHA-T1_UNSTRUCTURED_tf}
  \end{subfigure}
  \caption{Comparison of the phase field fracture solutions on unstructured meshes with the crack initialization techniques GEO-T1-WHL and PHA-T1. The top part shows the force-displacement curves along with the reference solution from the sharp crack model. The fields on the lower part show the phase field before and after propagation for both techniques with blue being uncracked ($\alpha=0$) and red being fully cracked ($\alpha=1$).}
  \label{fig:initial_cracks_in_unstructured_meshes}
\end{figure}

The crack initialization techniques GEO-T1-WHL and PHA-T1 properly represent an initial crack on a structured mesh.
Now, we want to check whether this result extends to unstructured meshes.
\Cref{fig:initial_cracks_in_unstructured_meshes} shows the results obtained with unstructured mesh.
We observe that the force-displacement curves in \cref{fig:Fu_unstructured} perfectly superimpose.
While more oscillations in the post-peak can be observed, the peak force remains mostly unchanged.
Note that the oscillations are due to the irregular mesh preventing the crack propagation in a perfect straight line, as shown in \cref{fig:crack_field_GEO-T1-WHL_UNSTRUCTURED_tf,fig:crack_field_PHA-T1_UNSTRUCTURED_tf}.
Those oscillations decrease with mesh refinement \parencite{loiseau_path-following_2025}.

\section{Discussion on the extensions to more complex cases}
\label{sec:discussion}

This study only tackles the case of rectilinear initial cracks and phase field models not relying on a history variable.
Extensions to curved cracks and models governed by a history variable can be envisaged and are discussed in the following paragraphs.

The extension of GEO-T1-WHL and PHA-T1 to curved initial cracks is primarily a technical difficulty.
For the geometric technique, the crack boundaries (either internal or external) must align with the crack path, and the crack itself must remain one-element wide.
For the phase field initial crack PHA-T1, the nodal values of the phase field must be set to $\alpha(\xx) = 1$ on all the nodes of elements intersecting with the crack path.
Ideally, the mesh nodes should be aligned with the initial crack so that the crack path passes through the centers of these elements.

For a phase field model governed by a history variable, such as the model proposed by \textcite{miehe_phase_2010}, all the proposed techniques can be applied.
However, the initial phase field can also be indirectly imposed through the history variable as it drives the crack phase growth.
As this method does not apply to every phase field model, it has not been studied in this study.
However, it corresponds to indirectly setting the initial crack phase through the history variable.
The outcome should be similar to that achieved using the initial phase field approach (PHA).

\section{Conclusion}
\label{sec:conclusion}

This study assesses various techniques for incorporating initial cracks in variational phase field fracture simulations so that it provides a good approximation of Griffith theory.
Different crack initialization techniques for phase field fracture simulations are compared to a sharp crack model, towards which the phase field model should theoretically $\Gamma$-converges.
This study shows that representing initial cracks as infinitely thin, for instance by duplicating nodes, leads to invalid propagation thresholds in phase field simulations.
Indeed, this approach induces peaks in the force-displacement curves and the critical load is significantly overestimated.
This conclusion is expected as, for phase field fracture simulations using continuous finite elements, cracks must be at least one element wide to properly represent the displacement jump across the crack.
Therefore, any initial crack must also be at least one element wide to maintain consistency and to avoid the artificial excess of energy required to transition to a one-element-wide zone.
Moreover, this studies shows that the crack phase field must be set to 1 (fully cracked state) along the entire initial crack boundary to avoid artificial toughening effects.

This study enables us to identify two unbiased techniques to incorporate initial cracks in phase field fracture simulation:
\begin{itemize}
  \item Geometric technique (GEO-T1-WHL): This approach consists of embedding a one-element-wide crack in the geometry (\emph{i.e.}, in the mesh) and imposing the crack phase field to the fully cracked state ($\alpha=1$) on it,
  \item Initial phase field technique (PHA-T1): This method initialized the phase field to the fully cracked state on the elements that the initial crack crosses.
\end{itemize}
Both techniques give equivalent results for a similar numerical cost, so both methods can be employed indifferently.

\section*{Backmatter information}

\paragraph{Funding}
The work was supported by Agence de l'Innovation de Défense -- AID -- via Centre Interdisciplinaire d'Etudes pour la Défense et la Sécurité -- CIEDS -- (projects 2022 - FracAddi), and by Agence Nationale de la Recherche (ANR-23-CE51-0054 3FAM).


\paragraph{Authors' contributions}
  F. Loiseau and V. Lazarus contributed equally to the conceptualization and methodology of the study.
  F. Loiseau was responsible for data collection, initial data analysis, and drafting the original manuscript.
  Both authors contributed to data interpretation and manuscript revision.
  V. Lazarus supervised the research and secured funding.
  Both authors reviewed and approved the final version of the manuscript.

\paragraph{Competing interests}
The authors declare that they have no competing interests.

\paragraph{Open Access} This article is licensed under a Creative Commons Attribution 4.0 International License, which permits use, sharing, adaptation, distribution and reproduction in any medium or format, as long as you give appropriate credit to the original author(s) and the source, provide a link to the Creative Commons license, and indicate if changes were made. The images or other third party material in this article are included in the article's Creative Commons license, unless indicated otherwise in a credit line to the material. If material is not included in the article's Creative Commons license and your intended use is not permitted by statutory regulation or exceeds the permitted use, you will need to obtain permission directly from the copyright holder. To view a full copy of this license, visit http://creativecommons.org/licenses/by/4.0/.

\printbibliography


\end{document}